\documentclass[%
 reprint,
 amsmath,amssymb,
 aps,
]{revtex4-2}

\usepackage{graphicx}
\usepackage{dcolumn}
\usepackage{bm}
\usepackage[usenames,dvipsnames]{xcolor} 
\newcommand{\Rone}{\textcolor{black}}
\newcommand{\Rtwo}{\textcolor{black}}

\begin{document}

\preprint{APS/123-QED}

\title{\Rtwo{Consolidation of freshly deposited cohesive and non-cohesive sediment: particle-resolved simulations}}

\author{Bernhard Vowinckel}
 \email{vowinckel@engineering.ucsb.edu}
\author{Edward Biegert}%
\author{Paolo Luzzatto-Fegiz}
\author{Eckart Meiburg}
\affiliation{%
 Department of Mechanical Engineering, University of California, Santa Barbara, CA 93106, USA
}%



\date{\today}

\begin{abstract}
We analyze \Rtwo{the consolidation of freshly deposited cohesive} and non-cohesive sediment by means of particle-resolved direct Navier-Stokes simulations based on the Immersed Boundary Method. The computational model is parameterized by material properties and does not involve any arbitrary calibrations. We obtain the stress balance of the fluid-particle mixture from first principles and link it to the classical effective stress concept. The detailed datasets obtained from our simulations allow us to evaluate all terms of the derived stress balance. We compare the settling of cohesive sediment to its non-cohesive counterpart, which corresponds to the settling of the individual primary particles. The simulation results yield a complete parameterization of the Gibson equation, which has been the method of choice to analyze self-weight consolidation.
\end{abstract}

\maketitle


\section{\label{sec:introduction}Introduction}
Fine grained sediments interact via attractive electric forces, commonly referred to as van der Waals (vdW) forces \citep{hamaker1937,visser1989}, and adhesive forces due to extracellular polymeric substances (EPS) such as biofilms \citep{black2002,fang2016}. Cohesive sediment thus behaves very differently from its cohesionless granular counterpart. This is especially true for sediment mixtures containing large quantities of cohesive sediment and organic matter, which are also known as `mud' \citep{amos1992,berlamont1993,sanford2001}. Mud deposition can lead to siltation of marine and riverine infrastructure or it can bind contaminants, which has important implications for ecology, sedimentology, and civil engineering as it is ubiquitous in various aquatic environments such as lakes, estuaries, and benthic habitats \citep{burchard2017,deswart2009}. The consolidation of mud is also important in the context of deep-sea hydrocarbon exploration \citep{meiburg2010}.

The deposition of mud can be subdivided into two processes: hindered settling \citep{kynch1952,richardson1954,ham1988,winterwerp2002,dorrell2010,dorrell2011} and consolidation \citep{been1981,townsend1990,sills1998,toorman1999,chauchat2013,zhou2016}. Both processes can happen simultaneously in the water column as sediment is still in the process of settling  until all suspended grains have made contact with the sediment bed that is supported by a bottom wall \citep{pane1985,toorman1996,winterwerp2002}. Consolidation is characterized by a contracting sediment bed due to the weight of the overlying deposits. This yields an excess pore pressure, which imposes an upward counterflow through the porous bed that is governed by the bed's permeability and the effective stress of the sediment. These processes have been the basis of the Gibson equation \citep{gibson1967}
\begin{equation}\label{eq:gibson_model}
  \frac{\partial \phi_v}{\partial t}=\frac{\partial }{\partial y}\left(\frac{k (\rho_p - \rho_f) \phi_v^2}{\rho_f} + \frac{k \phi_v}{\rho g}\frac{\partial \sigma_\text{eff}}{\partial y}\right)\qquad, 
\end{equation}
which is derived from one-dimensional mass and momentum conservation principles for the fluid and the sediment to predict the change of the horizontally averaged volume fraction $\phi_v$ of consolidating soils over time \citep{toorman1996}.  Here, $t$ is time, $y$ is the vertical coordinate, $\rho_p$ and $\rho_f$ are the particle and fluid densities,  $\rho= \phi_p \rho_p + (1-\phi_p) \rho_f$ is the averaged density of the suspension in a given control volume,  $g$ is the gravitational acceleration, $k$ is the permeability of the porous medium in units of length per time, and $\sigma_\text{eff}$ is the effective normal stress component in the $y$-direction. 

The effective stress has classically been assessed by means of the `effective stress concept' \citep{terzaghi1951}, which states that the total stress is the sum of the pore pressure and the effective stress. While it is a common definition that the effective stress reflects the part of the sediment weight that is supported by inter-particle contact, the experimental assessment of this physical quantity has been subject to debate. \citet{winterwerp2004} argued that this term is used as a mathematical concept to close the stress balance and \citet{sills1998} even claimed that there is no physical meaning to effective stress. Hence, while simulations using the Gibson equation \citep{lehir2011,grasso2015} show excellent agreement with experimental data, the model lacks general predictive capabilities since the empirical fitting of the parameters entering the Gibson equation has to be performed separately for each new experiment.

There are several reasons for the limitations of the Gibson model.  First, the effective stress concept remains one-dimensional, whereas it has been acknowledged that viscous fingering due to spatially heterogeneous particle concentrations can trigger instabilities that lead to complex three-dimensional flow features \citep{weiland1984,xu2016}. Furthermore, it was pointed out that every experimental setup inevitably contains sidewalls \cite{sills1998}. These sidewalls can lead to preferential drainage that introduces artifacts in the analysis. In recent years, experimental efforts have focused on the settling of sand-mud mixtures \citep{torfs1996,cuthbertson2008,manning2010,spearman2011,teslaa2013,teslaa2015}, where particles are too large to exhibit Brownian motion. Since the Gibson theory was derived for pure muds in the first place, applying the model to sand-mud mixtures might be problematic \citep{cuthbertson2016}. At the same time, the larger sized particles render the physical configuration attractive for phase-resolved simulations that allow for a full description of the fluid and the particle motion \citep{vowinckel2014,vowinckel2016,vowinckel2017a,vowinckel2017b,biegert2017,biegert2017peps}. 

The present study addresses these issues from a computational perspective. Following our earlier particle-resolved simulations \citep{vowinckel2018,biegert2018a,biegert2018b}, we present simulation data of settling cohesive and noncohesive sediment where we compute the motion of every particle in a fully resolved three-dimensional flow field without side walls. We present a detailed stress balance for the fluid-particle mixture that allows for a direct transfer of the governing equations to the classical `effective stress concept'. As a result, we provide a way to parameterize the Gibson equation in a straightforward fashion and illustrate the effects of cohesive forces on macroscopic particles by means of intergranular stresses in the sediment packing.

The paper is structured as follows: First, we briefly review our fully coupled computational approach to simulate particles in a viscous flow in section \ref{sec:method}. Then, we derive the 'effective  stress concept' from our governing equations in section \ref{sec:stress_budget}, before we present results for the settling behavior of cohesive grains of silt size and its noncohesive counterpart in section \ref{sec:results}. 

\section{Computational Approach}\label{sec:method}
\subsection{Particle-resolved Simulations}\label{sec:IBM}

We solve the unsteady Navier-Stokes equations for an incompressible Newtonian fluid, given by

\begin{equation} \label{eq:navier_stokes}
\frac{\partial{\textbf{u}}}{\partial{t}}+\nabla\cdot(\textbf{u}\textbf{u}) = -\frac{1}{\rho_f}\:\nabla p + \nu_f \nabla^2 \textbf{u} + \textbf{f}_\textit{IBM} \hspace{0.5cm},
\end{equation}
along with the continuity equation
\begin{equation}\label{eq:continuity}
\nabla\cdot\textbf{u}=0 \qquad 
\end{equation}
on a uniform rectangular grid with grid cell size $\Delta x = \Delta y = \Delta z = h$ following the scheme of \cite{biegert2017}. Here, bold and italic symbols represent vectors and scalar quantities, respectively, where $\textbf{u}=(u,v,w)^{T}$ designates the fluid velocity vector in Cartesian components, $p$ denotes pressure with the hydrostatic component subtracted out, $\nu_f$ is the kinematic viscosity, and $\textbf{f}_\textit{IBM}$ represents a volume force introduced by the Immersed Boundary Method (IBM) \citep{uhlmann2005,kempe2012a}.  This volume force acts in the vicinity of the inter-phase boundaries and couples the fluid phase to the particle motion. 

Within the framework of the IBM, we consider spherical particles as an approximation of primary particles from the grain-size fraction of silt. We calculate the motion of each individual spherical particle by solving an ordinary differential equation for its translational velocity $\textbf{u}_p=(u_p,v_p,w_p)^{T}$
       \begin{equation}\label{eq:part_trans}
        m_p\: \frac{\text{d}\textbf{u}_p}{\text{d} t} = \underbrace{\oint_{\Gamma_p} \boldsymbol{\tau} \cdot \textbf{n}\: {\text{d}A}}_{=\textbf{F}_{h,p}} +
        \underbrace{V_p\:( \rho_p-\rho_f )\: \textbf{g}}_{=\textbf{F}_{g,p}} + \textbf{F}_{c,p} \qquad ,
       \end{equation}
        and its angular velocity $\boldsymbol{\omega}_p=(\omega_{p,x},\omega_{p,y},\omega_{p,z})^{T}$
       \begin{equation}\label{eq:part_ang}
       I_p \:\frac{ \text{d}\boldsymbol{\omega}_p}{\text{d} t} = \underbrace{\oint_{\Gamma_p} \textbf{r}\times(\boldsymbol{\tau}\cdot\textbf{n})\:{\text{d}A}}_{=\textbf{T}_{h,p}} + \textbf{T}_{c,p} \hspace{0.5cm}.
      \end{equation}
Here, $m_p$ is the particle mass, $\Gamma_p$ the fluid-particle interface, $\boldsymbol{\tau}$ the hydrodynamic stress tensor, $\rho_p$ the particle density, $V_p$ the particle volume, $g$ the gravitational acceleration, $I_p=8\pi\rho_p R_p^{5}/15$ the moment of inertia, and $R_p$ the particle radius. Furthermore, the vector $\textbf{n}$ is the outward-pointing normal on the interface $\Gamma_p$, $\textbf{r} = \textbf{x} - \textbf{x}_p$ is the position vector of the surface point with respect to the center of mass $\textbf{x}_p$ of a particle, and $ \textbf{F}_{c,p}$ and $\textbf{T}_{c,p}$ are the force and torque due to particle interactions, respectively, which are computed using the Discrete Element Method (DEM) as described by \cite{biegert2017}. Furthermore, note the designation of the hydrodynamic force and torque as $\textbf{F}_{h,p}$ and $\textbf{T}_{h,p}$, respectively, as well as $\textbf{F}_{g,p}$ the force due to gravity.

We employ the approach of \cite{kempe2012a} for evaluating the IBM forces and solve \eqref{eq:part_trans} and \eqref{eq:part_ang} according to \cite{biegert2017}. The particles are explicitly coupled to the fluid motion through the hydrodynamic stress tensor $\boldsymbol{\tau}$ comprising viscous and pressure drag as a direct result of the IBM. The integration scheme subdivides the fluid time step into a total of 15 substeps to integrate \eqref{eq:part_trans} and \eqref{eq:part_ang} in time. It was shown by \citet{biegert2017} that this is necessary to resolve short-range particle-particle interactions such as lubrication and cohesive forces. 

Our simulation approach was validated by \cite{biegert2017} for the fluid-particle coupling of the method against experimental data of a sphere settling in an unbounded quiescent fluid \citep{mordant2000} as well as towards a wall \citep{tencate2002}. The particle-contact model was validated against benchmark data of \cite{gondret2002} and \cite{foerster1994}. The collective motion of a sediment bed sheared by a viscous flow was compared to the experimental data of \cite{aussillous2013} and we found satisfactory agreement.

\subsection{Particle-particle Interaction}\label{sec:collision}
We use the computational approach of \citet{biegert2017} for modeling cohesionless particle-particle interactions. This reference provides validation results for various benchmark experiments. In addition, the present study employs the cohesive force model proposed and validated by \citet{vowinckel2018}. The particle-particle interaction comprises short-range effects due to unresolved hydrodynamic lubrication forces $\textbf{F}_{l}$ and cohesive forces $\textbf{F}_\text{coh}$, as well as direct contact forces $\textbf{F}_d=\textbf{F}_n + \textbf{F}_t$ acting in the normal and tangential directions, denoted as $\textbf{F}_{n}$ and $\textbf{F}_{t}$. The resulting force on particle $p$ is the sum of all these effects
\begin{eqnarray}\label{eq:particle_forces}
	\textbf{F}_{c,p} = \sum_{q,\:q \neq p}^{N_p} \left( \textbf{F}_{l,pq} + \textbf{F}_{d,pq} + \textbf{F}_{\text{coh},pq}\right) + \nonumber\\
 \textbf{F}_{l,pw} + \textbf{F}_{d,pw} + \textbf{F}_{\text{coh},pw}\hspace{0.5cm},
\end{eqnarray}
where the subscripts $pq$ and $pw$ indicate interactions with particle $q$ or a wall, respectively. Detailed information about how we model the lubrication and direct contact forces is provided in \cite{biegert2017}. For the present study, we have chosen the same parameterization determined for silicate materials in experiments and used in our previous simulations \citep{biegert2017, joseph2001, joseph2004, vowinckel2014, vowinckel2017a, vowinckel2017b, vowinckel2018}.

To account for cohesive forces, we take an approach that is consistent with the theory for colloids developed by Derjaguin-Landau-Verwey-Overbeek (DLVO) \citep{derjaguin1941,verwey1948}, which states that there are two dominant short-range forces that can be interpreted as opposing potentials surrounding particles with grain sizes in the micro- to nanometer range. On the one hand, there exists a repulsive force when equally charged surfaces are in close proximity. On the other hand, as one particle causes correlations in the fluctuating polarization of a nearby particle surface an attractive force is generated. The former effect is usually called the repulsive `double-layer' (DL) force, while the latter effect is commonly referred to as van-der-Waals (vdW) force. These forces become important for gap sizes $\zeta_0 < \zeta_n < \zeta_\infty$, where $\zeta_0$ defines the microscopic size of surface asperities and 
$\zeta_\infty$ is the distance for which these forces decay to zero \citep{israelachvili1992}. The repulsive DL force and the attractive vdW force due to polarization scale as $F_\text{rep} \propto e^{-\zeta_n}$ and  $F_\text{att} \propto \zeta_n^{-2}$, respectively. The superposition of the two potentials yields a net force as a function of the gap size $\zeta_n$. A model incorporating both effects has been proposed by \citet{pednekar2017}
\begin{equation}\label{eq:DLVO}
 F_\text{DLVO} =\underbrace{ A_R \, R_\text{eff} \exp{\left(-\frac{\zeta_n}{\kappa}\right)}}_{F_\text{rep}} - \underbrace{\frac{A_H R_\text{eff}}{12(\zeta_n^2 + \zeta_0^2)}}_{F_\text{att}} \qquad ,
\end{equation}
where $A_R$ is a repulsive force scale for $\zeta_n = 0$ as a measure of the particles' surface potential, $\kappa$ is the Debye length, $A_H$ is the Hamaker constant, and $\zeta_0$ is the surface roughness preventing $F_\text{att}$ from diverging to infinity for vanishing gap size. These four parameters need to be adjusted according to the physical system. 

To account for these effects in particle resolved simulations, \citet{vowinckel2018} proposed a model  with the following properties: (i) it decays to zero as the gap size goes to zero, (ii) it has a maximum at a gap width orders of magnitude smaller than the particle diameter, and (iii) it decays to zero for larger gap sizes, without any discontinuous jumps. The physical idea of this model is described in  \cite{vowinckel2018}. The above properties are fulfilled with the mathematically simple model of a parabolic spring force
 \begin{equation}\label{eq:ansatz}
 \textbf{F}_\text{coh} = 
  \begin{cases}
  -k_\text{coh}(\zeta_n^2 - \zeta_n \lambda) \textbf{n}  & 0 < \zeta_n \leq \lambda \\
    0 & \text{otherwise} \qquad ,
  \end{cases} 
 \end{equation}
where $k_\text{coh} = \frac{A_H R_\text{eff}}{\zeta_0 \lambda^3 }$ denotes the stiffness constant, $\lambda$ represents the range over which the cohesive force is distributed, $A_H$ is the Hamaker constant, $\zeta_0$ is the minimal separation distance \citep{Israelachvili1974} and $R_\text{eff} = R_p R_q/(R_p + R_q)$ is the effective radius. The cohesive range $\lambda$ can be interpreted as a Debye length. 

The dimensional form \eqref{eq:ansatz} still requires the proper parameterization of the empirical parameters $A_H$ and $\zeta_0$. However, we can replace these empirical constants by writing \eqref{eq:ansatz} with respect to the maximum cohesive force $\text{max}(|| \textbf{F}_{\text{coh},50} ||)$. Choosing the median grain diameter $D_{50}$, the buoyancy velocity $u_s = \sqrt{g' D_{50}}$, the characteristic time scale $\tau_s = D_{50} / u_s$ and the characteristic mass $m_{50} = \rho_f \pi D_{50}^3/6$, the characteristic force scale for particles settling under gravity in an otherwise quiescent fluid becomes the specific weight  $m_{50} g'$, where $g' = (\rho_p - \rho_f) g/ \rho_f$ denotes the reduced gravity. After normalizing \eqref{eq:ansatz} with the specific weight we obtain
\begin{equation}\label{eq:cohesive_forces_dimensionless}
   \textbf{F}_\text{coh} =
  \begin{cases}
  - \text{Co} \, \frac{8 \, R_\text{eff}}{\lambda^2}(\zeta_n^2 - \zeta_n \lambda) \textbf{n}   & 0 < \zeta_n \leq \lambda \\ 
    0 & \text{otherwise} \qquad .
  \end{cases} 
\end{equation}
Hence, the characteristic parameter to define cohesive forces becomes the cohesive number
\begin{equation}\label{eq:cohesive_number}
 \text{Co} = \frac{\text{max}(|| \textbf{F}_{\text{coh},50} ||)}{m_{50} g'} \qquad .
\end{equation}
It represents the ratio of the maximum cohesive force for particles of diameter $D_{50}$ to the characteristic gravitational force scale of the problem \citep{sun2018}. 

\begin{figure}[t]
\centering
\includegraphics[width=0.5\textwidth]{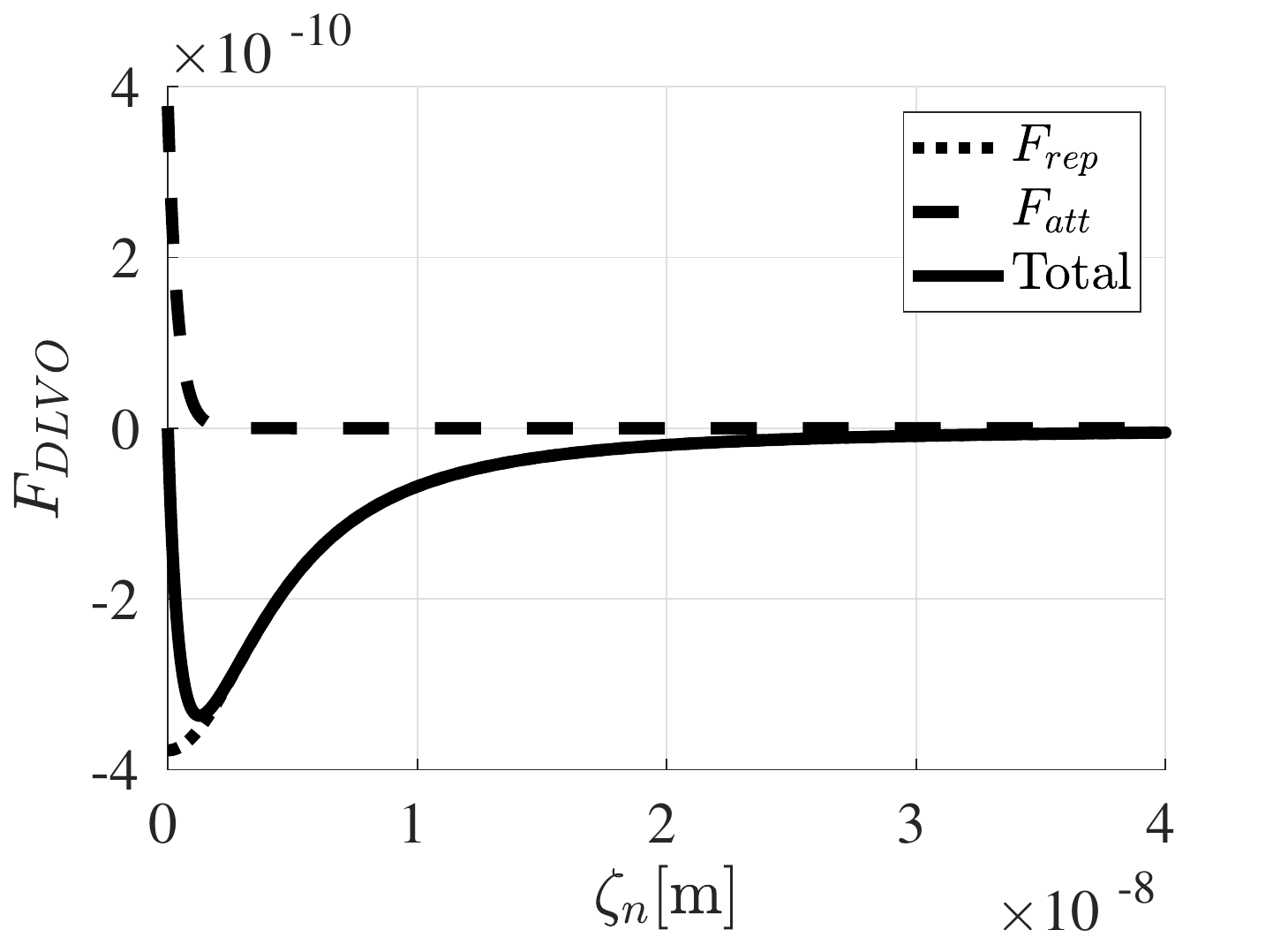}
        \caption{DLVO curve for $D_p = 20 \mu$m, $A_H = 1\cdot10^{-20}$,  $\zeta_0 = 4.7$nm, and $C_\text{salt} = 35$ppt.}
    \label{fig:DLVO_transfer}
\end{figure}

We can transfer the DLVO-theory \eqref{eq:DLVO} to our simpler model \eqref{eq:cohesive_forces_dimensionless} by choosing the following parameters: (i) $A_H = 1\cdot10^{-20}$ J, which reflects silica materials in water according to \cite{bergstrom1997}, (ii) $R_\text{eff} = \frac{R_p R_q}{R_p + R_q} = \frac{R_p}{2} = 5 \mu$m for monodisperse silt particles of grain size $D_p = 20 \mu$m, (iii) $\zeta_0 = 4.7$nm; (iv) we determine $\kappa$ using the approximation for the  monovalent salt sodium chloride given by \cite{berg2010} as $\kappa = \frac{0.304\cdot 10^{-9}\text{m}^{-1}}{|z| \sqrt{C_\text{salt}}}$ in meters, where $z$ is the valency of the salt and $C_\text{salt}$ is the salt concentration in mol/liter. Here, we choose the salinity of sea water with 35ppt. These parameters yield $C_\text{salt} = 0.6$mol/liter and $\kappa = 0.393$nm; (v) Since a key feature of our model is to have vanishing forces for particle contact, i.e. $\zeta_n = 0$, we set $A_R = A_H/(12 \zeta_0^2)$. The DLVO curve for this case is displayed in figure \ref{fig:DLVO_transfer}. In this scenario, the total force $F_\text{DLVO}$ follows the attractive forces with a distinct minimum at $\zeta_n \approx 1$nm. The minimum force is $| \min(F_\text{DLVO})|= 3\cdot 10^{-10}$N, while the weight becomes $F_g = \pi \, g (\rho_p - \rho_f) D_p^3/6 = 6.78 \cdot 10^{-11}$N, where we set gravitational acceleration, particle density, and fluid density to be $g = 9.81 \text{m/s}^2$, $\rho_p = 2650 \text{kg/m}^3$, and $\rho_f = 1000 \text{kg/m}^3$, respectively. This yields a cohesive number of $\text{Co} =  | \min(F_\text{DLVO})|/F_g = 5.0$. We found our parabolic spring model to be a good approximation of the curve shown by the solid line in figure \ref{fig:DLVO_transfer}. 

\Rtwo{Consequently, the cohesive number used in our simulations corresponds to the properties of fine to medium sized grains of silt settling in salt water. Note that the present approach can easily be extended to different types of cohesive sediment given that the temporal discretization is able to resolve the parabolic spring model during particle-particle interaction. In such case, the parameter needed to determine the cohesive number is $| \min(F_\text{DLVO})|$, i.e. the maximum attractive force. To investigate biofilms, for example, one can no longer use the present analogy of the DLVO-theory. Instead, the maximum attractive force would have to be determined experimentally \cite{gerbersdorf2018}}.

\setlength{\unitlength}{1cm}

\subsection{Simulation Scenario}\label{sec:scenario}
\begin{figure*}
  \includegraphics[width=\textwidth,trim={0 11cm 0 12cm},clip]{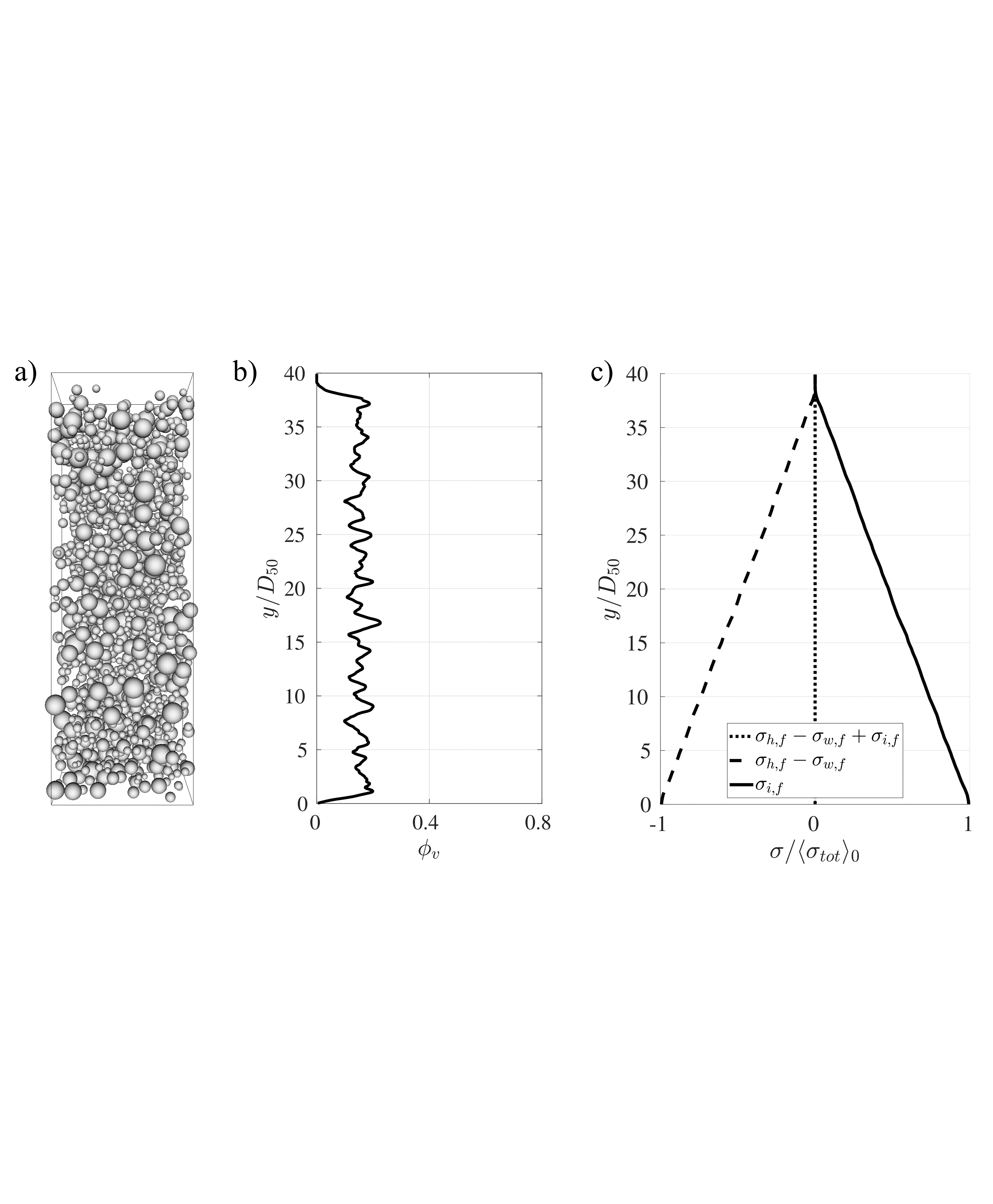}
        \caption{Initial settling behavior at $t = 17.6 \tau_s$. (a) Particle distribution, (b) particle volume fraction, and (c) vertical profiles of particle and fluid stress for $\text{Co}=0$ normalized by the total stress of the entire computational domain ($\Omega_{CV} = V_0$) according to \eqref{eq:eom_stress}.
        }
    \label{fig:initial_distribution}
\end{figure*}

To explore the influence of cohesive forces on the sedimentation process of a large, polydisperse ensemble of particles, we further analyze the simulation data first presented in \citet{vowinckel2018}, who investigated the hindered settling of silt particles. In this reference, a polydisperse mixture of particles with a relative density of $\rho_p / \rho_f = 2.6$ was placed in a tank of viscous fluid to obtain an initial volume fraction of $\phi_v = V_s / V_0 = 0.155$ (Figure \ref{fig:initial_distribution}a), where $V_s$ denotes the volume occupied by the particles and $V_0 = L_x \times L_y \times L_z = 13.1 D_{50} \times 40.0 D_{50}\times 13.1 D_{50}$ is the computational domain size. 

Consistent with the experiments of \cite{teslaa2015}, we chose a Reynolds number of $\text{Re} = u_s D_{50} / \nu_f = 1.35$. The size distribution of the polydisperse particles was continuous with a ratio  of ${\max\{D\}/\min\{D\} = 4}$ obeying a log-normal size distribution, \Rone{ which yields $\min\{D\}/D_{50}=0.6$ and $\max\{D\}/D_{50}=2.4$ for the smallest and largest particle, respectively}. A total of 1261 particles was placed randomly to obtain an almost uniform initial profile of $\phi_v$ (Figure \ref{fig:initial_distribution}b). \Rone{We conducted preliminary tests of dry settling, i.e. neglecting fluid forces, with varying particle distributions and domain sizes to investigate the dependency of the final deposit on the initial particle distribution. With the present domain being $40D_{50}$ in height, we found no dependence of the final configuration on the initial particle distribution, either.} We impose a no-slip condition at the bottom wall ($y=0$) and at the particle surfaces as well as a free-slip condition at the top wall ($y=L_y$), along with periodic boundary conditions in the wall-parallel $x$- and $z$-directions. This means that sidewalls are absent in our numerical simulations. Gravity is pointing towards the bottom wall in the  negative $y$-direction. \Rone{The minimum, median and maximum primary particle sizes are discretized by $\min\{D\}/h = 11$, $D_{50}/h = 18.25$ and $\max\{D\}/h = 44$ grid cells, respectively. It was shown by \citet{vowinckel2018} by a comparison to the analytical  particle  settling velocity that the grid resolution of our computational domain is fine enough to capture the settling behavior of all particle sizes at the particle Reynolds numbers encountered in the simulations presented in section IV}. We spread the cohesive forces over a shell of thickness $\lambda = h$, but it was shown by \citet{vowinckel2018} that the simulation results are not sensitive to this choice provided that $\lambda < R_p$. 
 \setlength{\unitlength}{1cm}

Two simulations were performed for different values of the cohesive number: (i) cohesionless grains with $\text{Co} =  0$, and (ii) cohesive sediment with $\text{Co} = 5$. For both simulations, \Rone{the particles with identical initial particle distributions were released from rest in a quiescent fluid to guarantee a straightforward comparison of (i) and (ii). Subsequently, the particles settle under the influence of gravity undergoing different settling behavior due to cohesive forces}. Following the scaling argument of \cite{pednekar2017} \Rtwo{presented in section IIB above, Co=5 corresponds to the properties of fine to medium sized silt in a saline ambient, i.e. the settling of macroscopic silica particles in ocean water}. For example, choosing a median grain size of $D_{50}=25 \mu$m renders the simulation domain 1mm tall. \Rone{As will be shown in section IV below, this choice allows for a simulation domain that is large enough to capture all relevant processes of particle settling so that the results can be evaluated along the lines of the `effective stress concept'.} 

\section{Stress Balance for the Fluid-particle Mixture}\label{sec:stress_budget}
\begin{figure}
\includegraphics[width=0.5\textwidth,trim={0 24cm 0 0},clip]{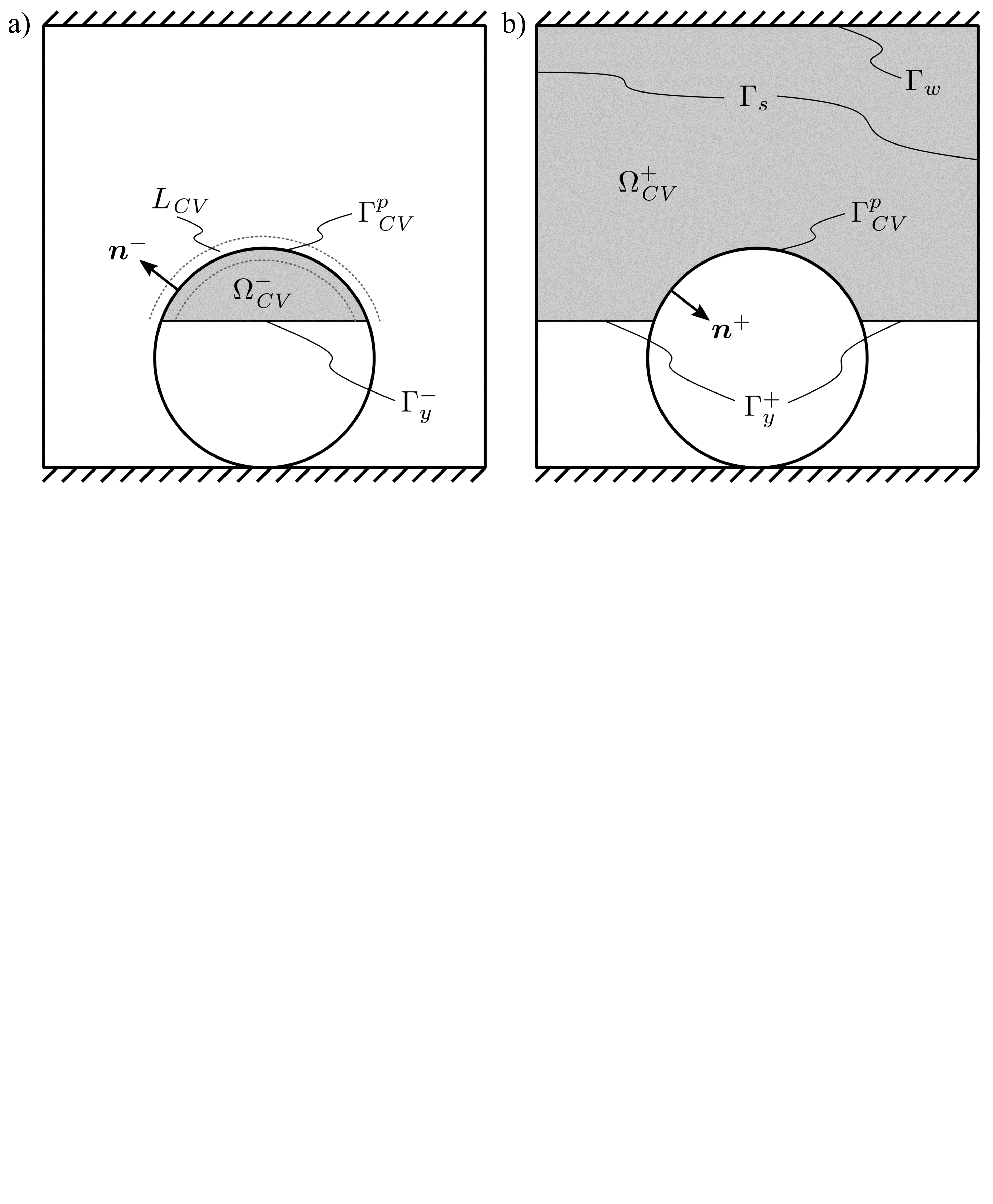}
        \caption{Schematic of the control volumes for (a) the particle interior and (b) the fluid surrounding the particle.}
    \label{fig:control_volume}
\end{figure}

To understand the settling and the consolidation of the fluid-particle mixture, we analyze the balance of the wall-normal stress components for the two phases separately. According to \citet{biegert2018a} and \citet{biegert2018b}, we can write the momentum balance of the fluid \eqref{eq:navier_stokes} in an integral sense to obtain fluid stresses for a control volume $\Omega_{CV}$ that extends from the top-wall ($y=L_y$) to an arbitrary height $y$ in the vertical direction and encompasses the entire domain in the $x$- and $z$-directions (figure \ref{fig:control_volume}). We can write the integral form of \eqref{eq:navier_stokes} as
\begin{equation} \label{eq:NS_int2}
\int\limits_{\Omega_\mathit{CV}^+} \rho_f \frac{\partial{\textbf{u}}}{\partial{t}} \,\text{d}V + \int\limits_{\Gamma_\mathit{CV}^+} \rho_f (\textbf{u}\textbf{u}) \cdot \textbf{n}^+ \,\text{d}A = \int\limits_{\Gamma_\mathit{CV}^+} \boldsymbol{\tau}^+ \cdot \textbf{n}^+ \,\text{d}A  ,
\end{equation}

where $\Gamma_\mathit{CV}^+ = \Gamma_w \cup \Gamma_s \cup \Gamma_y^+ \cup \Gamma_\mathit{CV}^p$ comprises all surfaces of the control volume shown for a single particle as a minimal example in Figure~\ref{fig:control_volume}b and $\textbf{n}^+$ is the normal vector pointing outwards from $\Omega_\mathit{CV}^+$. In addition, we recast the pressure and viscous terms using the fluid stress tensor, $\boldsymbol{\tau}^+ = -p \textbf{I} + \mu_f (\nabla \textbf{u} + (\nabla \textbf{u})^T)$, where $\textbf{I}$ is the identity matrix.  Here we neglect the effect of the immersed boundary force, assuming that it is implicitly handled by the fluid stress at the fluid/particle interface.  

\setlength{\unitlength}{1cm}

All of these terms except for the fluid stress at the particle surface are straightforward to calculate.  However, we can evaluate the fluid stress indirectly using the IBM force, as was done to obtain the particle equations of motion \eqref{eq:part_trans} and \eqref{eq:part_ang}.  That is, the IBM force acts as a jump in stress:
\begin{equation} \label{eq:IBM_CV_jump}
\int\limits_{L_\mathit{CV}} \textbf{f}_\text{IBM}\, {\text{d}V}
= \int\limits_{\Gamma_\mathit{CV}^p} \boldsymbol{\tau}^+ \cdot \textbf{n}^+\, {\text{d}A}
+ \int\limits_{\Gamma_\mathit{CV}^p} \boldsymbol{\tau}^- \cdot \textbf{n}^-\, {\text{d}A} ,
\end{equation}
where we are careful to distinguish between $\textbf{n}^+$, the outward surface normal for the volume $\Omega_\mathit{CV}^+$, and $\textbf{n}^-$, the outward surface normal for the volume $\Omega_\mathit{CV}^-$, which point in opposite directions.

To evaluate the fluid stress $\boldsymbol{\tau}^- \cdot \textbf{n}^-$ in the particle interior, we can perform a stress balance on $\Omega_\mathit{CV}^-$ (figure \ref{fig:control_volume}a). The integral form of the Navier-Stokes equations together with divergence theorem give us
\begin{equation} \label{eq:NS_in_int3}
\int\limits_{\Gamma_y^-} \rho_f (\textbf{u}\textbf{u}) \cdot \textbf{n}^- \,\text{d}A
= \int\limits_{\Gamma_\mathit{CV}^p} \boldsymbol{\tau}^- \cdot \textbf{n}^- \,\text{d}A
+ \int\limits_{\Gamma_y^-} \boldsymbol{\tau}^- \cdot \textbf{n}^- \,\text{d}A.
\end{equation}

Using \eqref{eq:NS_in_int3} together with \eqref{eq:IBM_CV_jump} and \eqref{eq:NS_int2}, we obtain
\begin{eqnarray} \label{eq:mom_fluid}
\underbrace{\int\limits_{\Omega_\mathit{CV}^+} \rho_f \frac{\partial{\textbf{u}}}{\partial{t}} \,\text{d}V }_\text{Acceleration term}
\underbrace{
\int\limits_{\Gamma_w} \boldsymbol{\tau}^+ \cdot \textbf{n}^+ \,\text{d}A
+ \int\limits_{\Omega_\mathit{CV}} \textbf{f}_b \,\text{d}V
}_\text{External force}
= \nonumber\\ \underbrace{
-\int\limits_{\Gamma_y} \boldsymbol{\tau} \cdot \textbf{n} \,\text{d}A
+ \int\limits_{\Gamma_y} \rho_f (\textbf{u}\textbf{u}) \cdot \textbf{n} \,\text{d}A
}_\text{Fluid force} 
\underbrace{
- \int\limits_{L_\mathit{CV}} \textbf{f}_\text{IBM}\, {\text{d}V}
}_\text{Particle force} ,
\end{eqnarray}
where $\Omega_\mathit{CV} = \Omega_\mathit{CV}^+ \cup \Omega_\mathit{CV}^-$, $\Gamma_w$ is the area of the top wall, and $\Gamma_y=\Gamma_y^+\cup \Gamma_y^-$.

Using the definition of the horizontal, superficial average \cite{nikora2013},
\begin{equation}\label{eq:average}
\left< \theta\lvert_y \right> = \frac{1}{L_x L_z} \int_0^{L_z} \int_0^{L_x} \theta(x,y,z,t) \,\text{d}x \, \text{d}z ,
\end{equation}
we can rewrite \eqref{eq:mom_fluid} for the $y$-velocity component as
\begin{align} \label{eq:fx_stress}
\underbrace{\frac{1}{A_w}\int\displaylimits_{\Omega_{CV}}  \rho_f \frac{\partial v}{\partial t} \text{d}V}_\text{Local acceleration}
+\underbrace{
\left< p\lvert_{L_y} \right> + 2\nu_f \rho_f \left<\left.\frac{\partial v}{\partial y}\right|_{L_y}\right>
}_\text{Stress on the top wall} = \nonumber \\ 
 \underbrace{
-\left< p\lvert_y \right> +
2\nu_f \rho_f\left<\left.\frac{\partial u}{\partial y} \right|_y \right>
- \rho_f \left<vv|_y \right>
}_\text{Fluid stress} \nonumber \\  
\underbrace{
- \int_y^{L_y} \left<f_{\mathit{IBM,x}}\right> {\text{d}y}
}_\text{Particle stress} ,
\end{align}
where $A_w = L_x L_z$ is the area of the bottom wall. For this derivation, we have used the fact that $\mu_f$ and $\rho_f$ are constant throughout the domain.

This yields the stress balance of the fluid phase for the horizontally averaged, vertical component of the normal stress
\begin{equation}\label{eq:stress_budget}
 \sigma_{t,f} =  \sigma_{h,f} - \sigma_{w,f} + \sigma_{i,f} \qquad .
\end{equation}
Here, 
\begin{equation}\label{eq:local_stress}
\sigma_{t,f}(y,t) = \frac{1}{A_w}\int\displaylimits_{\Omega_{CV}}  \rho_f \frac{\partial v}{\partial t} \text{d}V
\end{equation}
is the stress due to local acceleration. Further, 
\begin{equation}\label{eq:wall_stress}
\sigma_{w,f}(t) = -\langle p_w|_{L_y} \rangle + 2 \rho_f \nu_f \left\langle \left. \frac{\partial v}{\partial y}  \right \rvert_{L_y} \right \rangle
\end{equation}
is the hydrodynamic stress on the top wall.
The angular brackets denote horizontal averaging. The hydrodynamic stress due to fluid motion becomes 
\begin{equation}\label{eq:hydrodynamic_stress}
\sigma_{h,f}(y,t) = -\langle p |_y \rangle + 2\nu_f \rho_f \left\langle \left. \frac{\partial v}{\partial y}  \right \rvert_{y} \right \rangle - \rho_f \left \langle  vv |_y \right \rangle
\end{equation}
which comprises pressure and viscous forces as well as convection.  Finally,
\begin{equation}\label{eq:interfacial_stress}
\sigma_{i,f}(y,t) = -\frac{1}{A_w}\int\displaylimits_{S_{int}} \boldsymbol{\tau} \cdot \textbf{n}\: {\text{d}A}
\end{equation}
is the interfacial stress exerted on the fluid by the particles, where $S_{int}$ is the total area of the fluid particle interface enclosed in the control volume $\Omega_{CV}$. In our simulations, we found that $\sigma_{w,f} - \sigma_{h,f} = \sigma_{i,f}$ holds at all $y$-locations and all times (e.g. figure \ref{fig:initial_distribution}c). Furthermore, the fluid stress was dominated by the pressure term, so that we conclude $\sigma_{w,f} - \sigma_{h,f} \approx \langle p\lvert_y \rangle$, provided that viscous stresses and stresses due to convection are small. Since the hydrostatic component is subtracted out for \eqref{eq:navier_stokes}, $p$ is equivalent to the excess pressure in the water column \citep{winterwerp2004}.

For the particle phase, we sum over all particles within the control volume for the different terms in the particle equation of motion \eqref{eq:part_trans} to get integral quantities:
       \begin{align} \label{eq:eom_stress}
        \underbrace{\frac{1}{A_w}\sum_{N_{CV}} m_p\: \frac{\text{d}v_p}{\text{d} t}}_{=\langle\sigma_{t,p}\rangle_{y}} = \underbrace{\frac{1}{A_w}\sum_{N_{CV}} \oint_{\Gamma_p} \tau_{yy} \cdot n_{y}\: {\text{d}A}}_{=\langle\sigma_{i,p}\rangle_{y}=\langle p\rvert_{y} \rangle } + \nonumber \\
        \underbrace{\frac{1}{A_w}\sum_{N_{CV}} V_p\:( \rho_p-\rho_f )\: g_y}_{=\langle\sigma_\text{tot}\rangle_{y}} + \underbrace{\frac{1}{A_w}\sum_{N_{CV}}F_{c,p,y}}_{=\langle\sigma_\text{eff}\rangle_{y}} \qquad ,
       \end{align}
where $N_{CV}$ is the number of particles enclosed in $\Omega_{CV}$ and the operator $\langle\cdot\rangle_{y}$ indicates the volume average over $\Omega_{CV}$, which is to be distinguished from the horizontal averaging operator $\langle\cdot\rangle$ defined by equation \eqref{eq:average}. 

The explicit link between the two phases via the IBM yields that $\sigma_{i,f}$ is equal to the static excess pressure (cf. figure \ref{fig:initial_distribution}c). Hence, \eqref{eq:stress_budget} and \eqref{eq:eom_stress} are coupled through the equality $-\sigma_{i,f} = \langle p\rvert_{y} \rangle  = \langle\sigma_{i,p}\rangle_y$. Using this equality and assuming $\langle \sigma_{t,p}\rangle_y$ to be small, we can interpret \eqref{eq:eom_stress} in terms of the 'effective stress concept' \citep{terzaghi1951}. It was described in this reference that 
\begin{equation}\label{eq:eff_stress}
   -\langle\sigma_\text{tot}\rangle_{y} = \langle p\rvert_{y} \rangle+   \langle\sigma_\text{eff}\rangle_{y}    \qquad,
\end{equation}
where $-\langle\sigma_\text{tot}\rangle_y$ is the weight of the particles submerged within $\Omega_{CV}$, $\langle p\rvert_{y} \rangle$ is the pore water pressure at location $y$ and $\langle\sigma_\text{eff}\rangle_{y}$ is the effective stress due to particle interactions. For this reason, we normalize all stresses by the total submerged weight of all the particles $\langle \sigma_\text{tot} \rangle_0$ in the following section. 

\setlength{\unitlength}{1cm}
\begin{figure*}
\includegraphics[width=\textwidth]{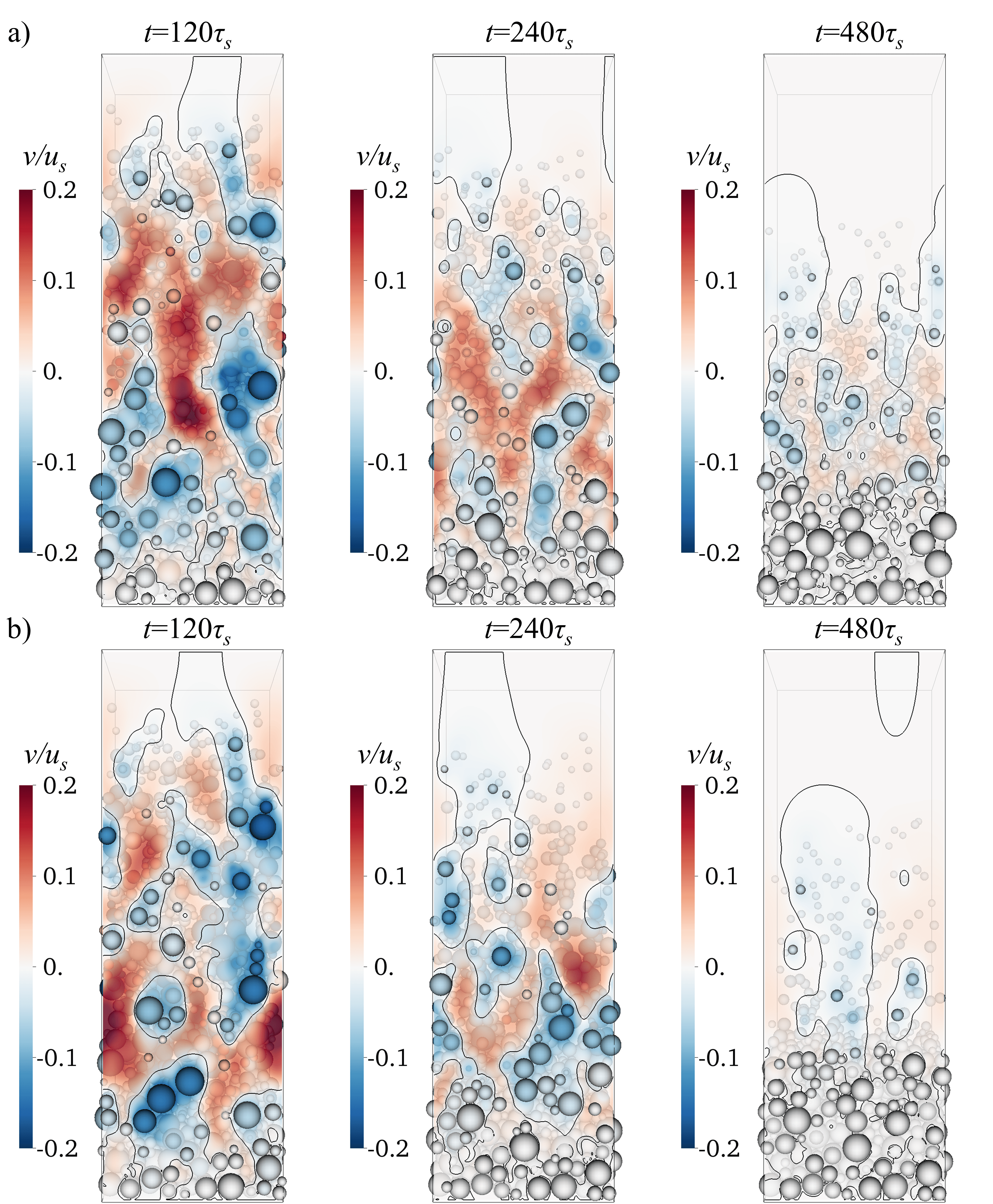}
        \caption{Particle configurations during the settling process. (a) $\text{Co}=0$ and (b) $\text{Co}=5$. The color scheme reflects the vertical velocity, and the solid black line denotes the contour of vanishing vertical fluid velocity.}
    \label{fig:settling}
\end{figure*}

The weight of the sediment can be computed in a straightforward manner:
\begin{equation}  \label{eq:weight}
     \langle \sigma_\text{tot}\rangle_{y} = (\rho_p - \rho_f) g \int_y^{L_y} \phi_v(y,t) \text{d}y  \qquad ,
\end{equation}
where $\phi_v$ is particle volume fraction, and $L_y$ is the length of the domain in $y$-direction. The pressure is computed as the horizontal average in plane $y$, which is the lower bound of the control volume under consideration:
\begin{equation}  \label{eq:pressure}
     \langle p\lvert_y\rangle = \frac{1}{A_w}\int_0^{L_z}\int_0^{L_x} p(x,y,z,t) \text{d}x \, \text{d}z   \qquad .
\end{equation}

To compute depth-resolved distributions of effective stresses, we introduce a horizontal averaging operator that uses a step function to distinguish between particles and fluid:
\begin{equation}\label{eq:CGM_new}
  \langle \sigma_\text{eff} \rangle_y = \frac{1}{A_w}\int_0^{L_z}\int_0^{L_x} \frac{\Delta V_p}{V_p} \, F_{c,p,y}(x,y,z,t) \, \text{d}x \, \text{d}z  \qquad ,
\end{equation}
where  $A_w$ is the area of the bottom wall and $\Delta V_p$ is the volume of particle $p$ cut by the horizontal slice, and $V_p$ is the total volume of this particle. This operator is very similar to the one used by \cite{vowinckel2018} to compute averages of particle velocities.

\section{Results}\label{sec:results}
\subsection{Settling Behavior}\label{sec:settling}

After the initial release of the particles from rest, the grains start to settle towards the bottom. As shown by \citet{vowinckel2018}, the kinetic energy of the particles peaks at $t = 17.6 \tau_s$. The stress balance for this moment is shown in Figure \ref{fig:initial_distribution}c.  Since both the cohesionless and the cohesive simulation are initialized identically, their results collapse for this initial stir-up phase. During this phase, particles have not yet settled out and all the static pressure induced by the particle weight is transferred to the fluid pressure. As the particle distribution is still fairly uniform  (figures \ref{fig:initial_distribution}a and b), the two profiles shown in figure \ref{fig:initial_distribution}c are approximately linear in the $y$-direction. The excess pore pressure builds up throughout the water column to reach a balance with $\langle \sigma_{tot}\rangle_0$ at the bottom wall. 

\Rone{Figure \ref{fig:settling} shows the settling process over time as snapshots taken at $t/\tau_s=25\% \, T_\text{sim}$, $t/\tau_s=50 \%  \, T_\text{sim}$ and $t/\tau_s=100\%  \, T_\text{sim}$, where $T_\text{sim}$ is the total simulation time. The figures show both, the instantaneous particle distribution and a translucent contour slice cutting through the front of the domain showing the vertical fluid velocity component. Particles are also colored by their wall-normal velocity component following the color scheme of the contour slices. Since particles appear as opaque objects, the grains in the front are  blocking the view on the particles in the back of the domain. Note, however, that since we are resolving the motion of each particle individually, the number of particles remains constant throughout the simulations.} 

\Rone{As particles start to settle, they replace fluid at the bottom of the tank and generate an upward counterflow (red regions in figure \ref{fig:settling} at $t=120 \tau_s$).} For the current simulations, this counterflow is sufficiently strong to sweep smaller particles upward. This effect is more pronounced for cohesionless particles. The upwelling fluid represents one mechanism for hindered settling and for the segregation of the grain sizes for very large water columns \citep{teslaa2015}. Indeed, it was shown by \citet{vowinckel2018} that the present data yields excellent agreement with the classical hindered settling functions of \citet{richardson1954} and \citet{winterwerp2002}.

\setlength{\unitlength}{1cm}
\begin{figure}[t]
\centering
\includegraphics[width=0.3\textwidth]{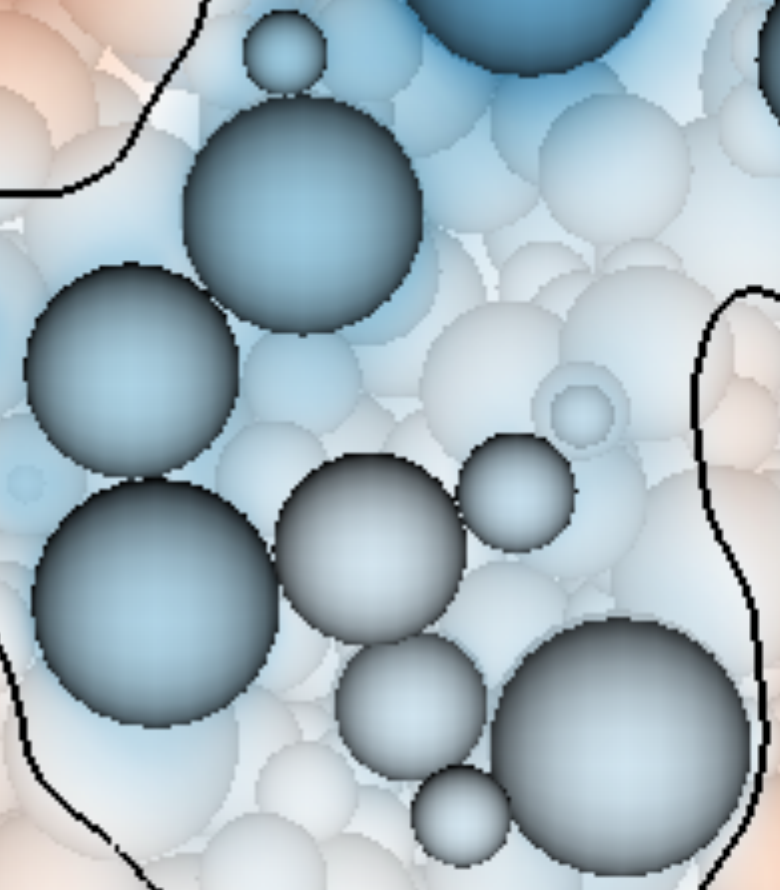}
        \caption{Zoom into the lower third of the domain with settling cohesive sediment (figure \ref{fig:settling}b at $t=240\tau_s$).}
    \label{fig:settling_zoom}
\end{figure}

As time progresses the impact of cohesive forces on the settling behavior becomes evident (figure \ref{fig:settling} at $t=240 \tau_s$). The cohesive sediment starts to built aggregates as particles attach to each other once they come into close enough contact. \Rtwo{The flocculation process is exemplified in figure \ref{fig:settling_zoom}, which shows a zoom into figure \ref{fig:settling}b at $t=240\tau_s$. For the current physical configuration, cohesive particles are able to form chains of several particles with varying diameter. } Hence, small cohesive particles that bond to bigger ones move with the settling velocity of the large particle.  As a result, smaller particles settle faster than individual primary particles would. This observation is in line with experimental evidence \citep{mehta1989,lick1993,winterwerp1998}. 

\begin{figure*}
\includegraphics[width=\textwidth,trim={0 0cm 0 0cm},clip]{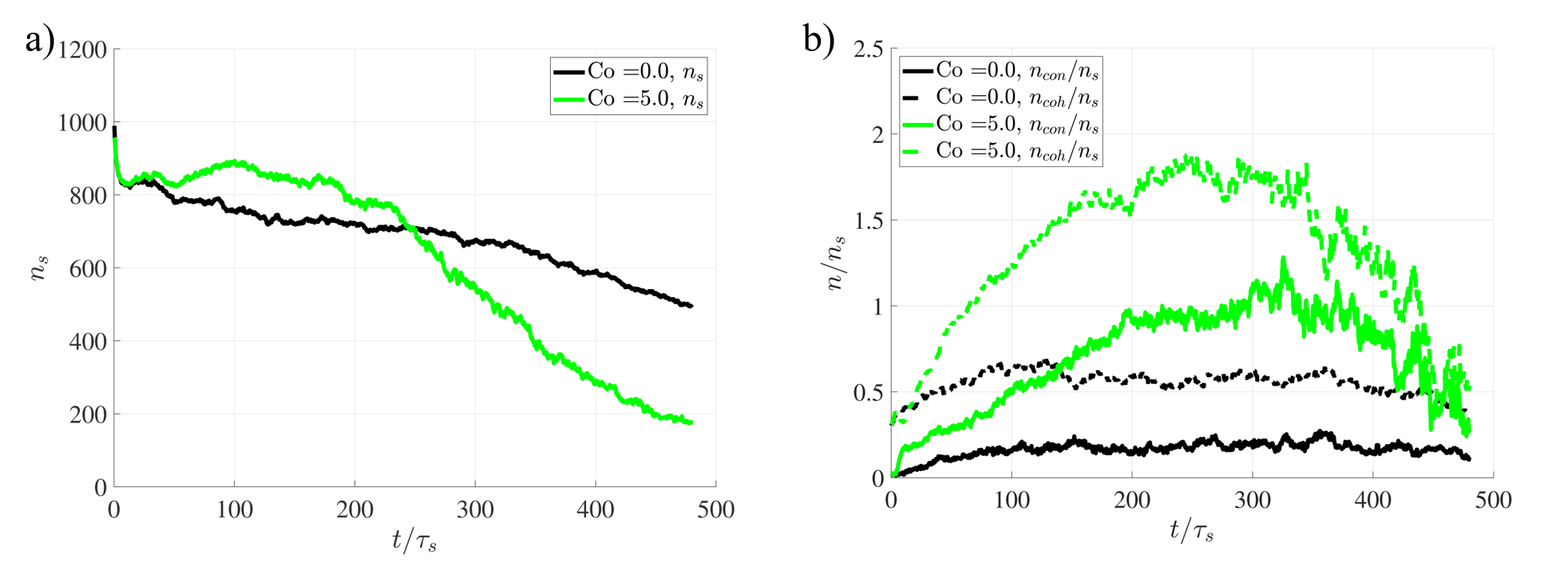}
        \caption{ a) Number of settling particles ($v_p > 0.01 u_s$) over time. b) Number of direct contacts $n_\text{con}$ and short-range interactions $n_\text{coh}$ for all settling particles. }
    \label{fig:number_interaction}
\end{figure*}

\Rone{This observation is addressed in a statistical sense in figure \ref{fig:number_interaction}. Figure \ref{fig:number_interaction} shows the number of settling particles $n_s$ (conditioned by $v_p > 0.01 u_s$). Initially, particles start to accelerate and those that were initially close to the bottom are immediately filtered out. Later in time, a steady decrease of $n_s$ over time can be observed for the cohesionless sediment, whereas for cohesive sediment $n_s$ shows a slight increase reaching a local maximum at $t\approx 100 \tau_s$. During this time, particles are moving rather fast through the domain, and once they get into contact, they form larger aggregates such as the one depicted in figure \ref{fig:settling_zoom}. Subsequently, the number of settling particles decreases more rapidly for cohesive sediments as the aggregates settle faster than the noncohesive sediments. }

\Rone{The aggregation process is further addressed by counting all particle-particle interactions of all settling particles. Throughout the entire simulation time, the average number of direct contacts $n_\text{con}$ and the average number of short-range interactions $n_{coh}$ (within the distance $\lambda$) remains fairly constant for cohesionless grains. Note that for this type of sediment, short-range interactions are influenced by lubrication forces only. Cohesive sediment, on the other hand, forms larger aggregates with up to three different  particles on average considering direct contacts and cohesive short-range interaction combined. As larger aggregates settle faster than individual particles, these aggregates make contact with the bottom wall earlier in time, so that the number of particle-particle interactions summed over all settling particles starts to decay at $t\approx 300 \tau_s$. } As the cohesive sediment settles out faster than the noncohesive sediment, the counterflow decays earlier in time (figure~\ref{fig:settling}b $t = 480 \tau_s$).

\setlength{\unitlength}{1cm}

\subsection{Transition from Hindered Settling to Consolidation}\label{sec:transition}
\begin{figure*}
\includegraphics[width=\textwidth,trim={0 0cm 0 0cm},clip]{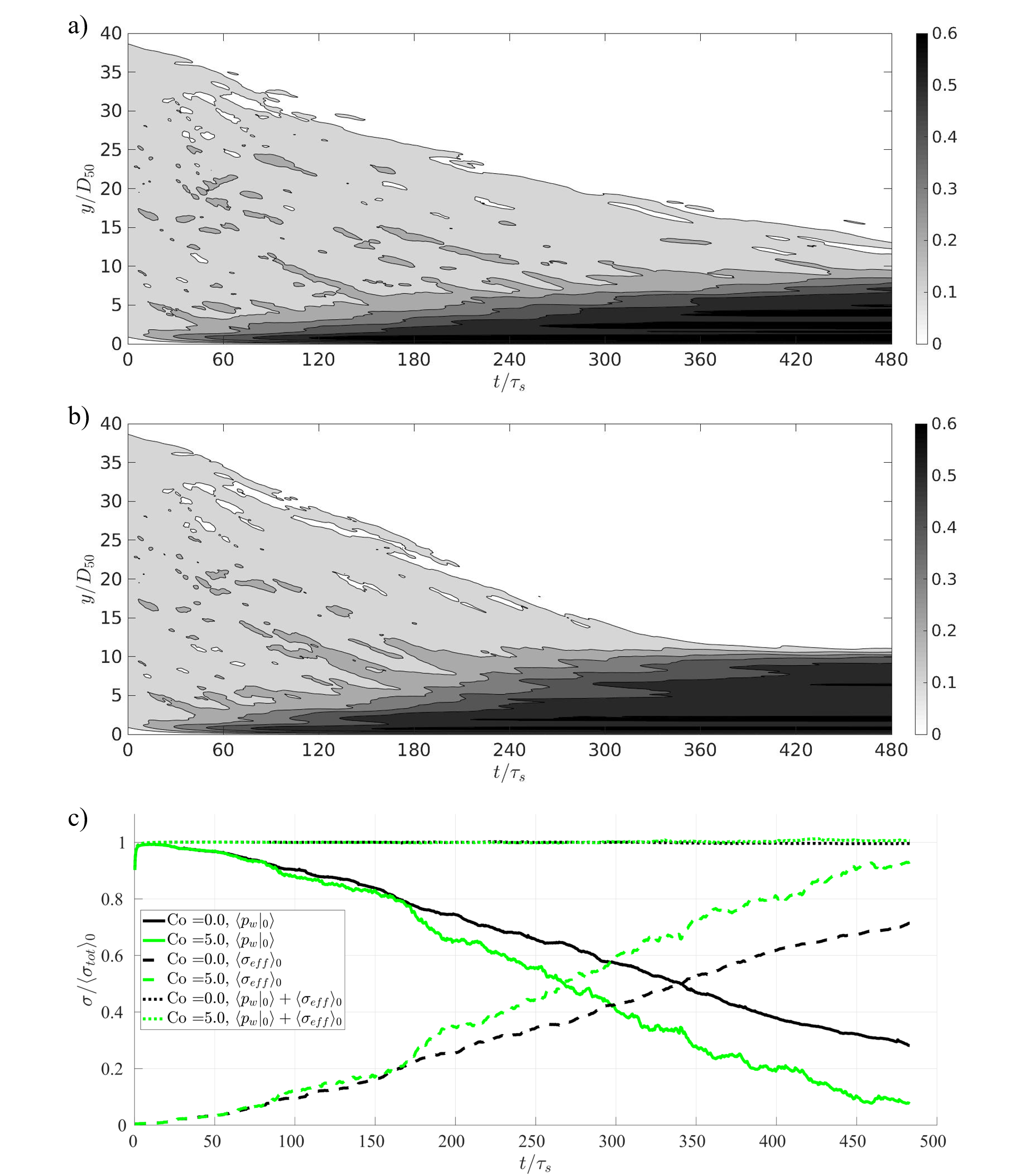}
        \caption{ Contours of the horizontally-averaged particle volume fraction $\phi_v$ over time: (a) cohesionless sediment and (b) cohesive sediment. (c) Time evolution of the different components of the particle stress balance acting on the bottom wall. }
    \label{fig:settling_concentration}
\end{figure*}

The enhanced settling speed for $\text{Co}=5$ is confirmed by the horizontally averaged concentration profiles plotted over time (Figure \ref{fig:settling_concentration}a and b) following the analysis of \citet{been1981}. Up to $t=120 \tau_s$, the volume fraction contours for the two simulations are nearly identical, as cohesive forces have not yet had sufficient time to cause a noticeable change. As time progresses, two fronts become clearly visible for both simulations. The upper front marks the transition between the clear fluid and the suspended sediment, whereas the second front shows the transition between the suspended sediment and the sediment bed. The isoline marking this front is called the gelling concentration \citep{winterwerp2002}. Since cohesive forces result in the formation of flocs with larger settling speeds, particles accumulate at the bottom of the tank more quickly. At $t\approx 350 \tau_s$, the two fronts merge into one for the simulation data of the cohesive sediment. This point in time is called the point of contraction (PoC)  and marks the transition between hindered settling and consolidation \citep{winterwerp2004}.  The cohesionless sediment, on the other hand, has not yet reached the PoC by the end of the simulation. 

\setlength{\unitlength}{1cm}

The transition from hindered settling to consolidation can also be analyzed in terms of the effective stress concept \eqref{eq:eff_stress}. To this end, we compute all stress components for $\Omega_{CV}=V_0$, i.e. the entire computational domain. \Rone{To evaluate the effective stress, we compute the normal stresses due to contact forces in the wall-normal direction. Note that we also computed the normal stress components in the periodic $x$- and $z$-directions as will be shown in section IV D below. We found these two wall-parallel components to be identical, which yields transversal isotropic conditions. This proves that the width of the simulation domain is large enough to capture all relevant effects of particle-particle interactions in these directions.} 

We plot all components entering the effective stress balance (equation \eqref{eq:eff_stress}) over time in figure \ref{fig:settling_concentration}c. It was shown in section \ref{sec:settling} that particles are deposited on the bottom more rapidly for cohesive sediment. These dynamics are also reflected in the stress balance of the fluid-particle mixture. Recall that $\langle p\rvert_0 \rangle$ is the fluid stress acting on the particle phase, which after the initial phase is equivalent to the weight of those particles that are still suspended \citep{been1981,toorman1996}. On the other hand, $\langle \sigma_\text{eff}\rangle_0$ captures all particle-particle interactions. Since the forces between two interacting particles in suspension are opposite and equal, these do not cause a net force on the bottom wall. Thus, $\langle \sigma_\text{eff} \rangle_0$ reflects the weight of those particles that are supported by the external contact forces with the bottom wall. After the initial increase of $\langle p\rvert_0 \rangle$ up to $\langle \sigma_\text{tot}\rangle_0$, when all of the particle weight is supported by fluid forces (cf. figure \ref{fig:initial_distribution}c), $\langle p\rvert_0 \rangle$ decays over time, whereas $\langle \sigma_\text{eff}\rangle_0$ increases at the same rate. Hence, less and less of the particle weight is supported by fluid forces, and more and more of it is supported by interparticle forces within the sediment bed, which illustrates the transition from hindered settling to consolidation \citep{winterwerp2004}. While $\langle p\rvert_0 \rangle$ and $\langle \sigma_\text{eff}\rangle_0$ behave very similarly for cohesive and noncohesive sediment until $t = 180 \tau_s$, cohesive sediment experiences a systematic shift at this time that illustrates the enhanced settling due to flocculation. Particles that assemble in flocs are seen to make contact with the wall at earlier times. Apart from the initial increase, we obtain $\langle p\rvert_0 \rangle + \langle \sigma_\text{eff}\rangle_0 \approx \langle \sigma_\text{tot}\rangle_0$, whereas the rate of change for both components is approximately linear. This suggests that the local acceleration does not add to the stress balance of the particle motion even though the processes under investigation are inherently transient. Interestingly, we obtain $\langle \sigma_\text{eff}\rangle_0=0.71 \langle \sigma_\text{tot}\rangle_0$ for the cohesive sediment at the PoC ($t \approx 350 \tau_s$). This is the same value observed for cohesionless sediment at the end of the simulation time $t=480\tau_s$, where the PoC has not been reached (figure \ref{fig:settling_concentration}a), but the height of the cohesive sediment bed is much thicker compared to the cohesionless sediment bed. Hence, comparing figures \ref{fig:settling_concentration}a and b, we conclude that flocculation promotes larger pore spaces due to different vertical distributions of effective stresses. 

\begin{figure*}
{\includegraphics[width=\textwidth,trim={0 0cm 0 0cm},clip]{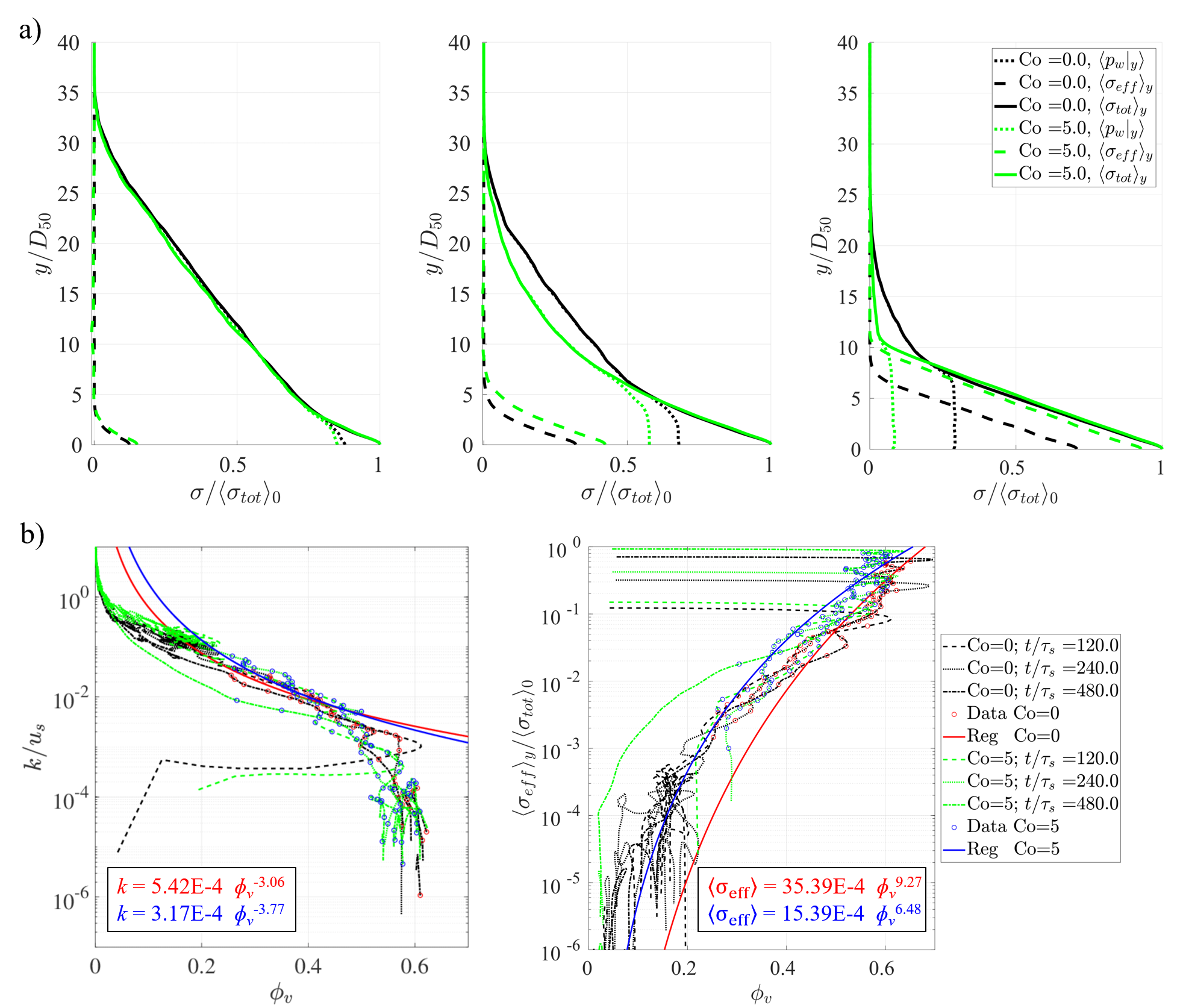}}
        \caption{(a) Vertical stress distributions reflecting the configurations illustrated in Figure \ref{fig:settling};  (b) Permeability $k$ and effective stress $\langle \sigma_\text{eff} \rangle_y$ as a function of particle volume fraction $\phi_v$ to parameterize the Gibson equation \eqref{eq:gibson_model}. Circles mark the data used to fit the regression function (`Reg') given in the figure.
        }
    \label{fig:fluid_stress}
\end{figure*}

While \eqref{eq:eom_stress} provides a measure for the volume averaged stresses, we can also evaluate the stress terms as vertical profiles for all components of \eqref{eq:eff_stress}. The instants shown in figure \ref{fig:fluid_stress}a correspond directly to the situations in figure \ref{fig:settling}.  As expected, we obtain $\langle p\rvert_y \rangle = \langle\sigma_\text{tot}\rangle_y$ far away from the wall, where all particles are fully supported by the fluid. Initially, $\langle p\rvert_y \rangle$ decreases linearly with height ($t=120\tau_s$). Whereas $\langle \sigma_\text{tot}\rangle_y$ decreases linearly in the suspended region, the profile for cohesive sediment becomes convex as flocs start to form, which indicates an accelerated settling process.  

The increase of the sediment bed thickness is shown by regions with $\langle\sigma_\text{eff}\rangle_y>0$. Again, $\langle\sigma_\text{eff}\rangle_y$ increases linearly with height. In addition, $\langle\sigma_\text{eff}\rangle_y$ and $\langle p\rvert_y \rangle$ add up to $\langle\sigma_\text{tot}\rangle_y$ by an out-of-balance of less than 1\%, illustrating the capability of our method to explicitly compute the stress balance of the entire fluid-particle mixture.

 \setlength{\unitlength}{1cm}

\subsection{Parameterization of the Gibson Model}\label{sec:gibson}
The Gibson equation \eqref{eq:gibson_model} was derived by \cite{toorman1996} for the simultaneous treatment of sedimentation and self-weight consolidation by using continuity principles and the Darcy-Gersevanov law:
\begin{equation}\label{eq:darcy}
 (1-\phi_v) (v_f - v_p) = -k \frac{1}{g \rho_f}\frac{\partial \langle p\lvert_y\rangle}{\partial y} \qquad ,
\end{equation}
where $v_f$ is the horizontally averaged vertical fluid velocity and $v_p$ is the horizontally averaged particle velocity. Solving \eqref{eq:gibson_model} requires constitutive relationships for the parameters $k$ and $\sigma_\text{eff}$ \citep{merckelbach2004}. According to \cite{toorman1996}, it has been a common assumption that $k$ and $\sigma_\text{eff}$ depend on the volume fraction only, even though it was discussed in this reference that the assumption can not hold for $\sigma_\text{eff}$, which must be zero in a suspension while $\phi_v$ is not. 

We can test this hypothesis by computing $k$ and $\sigma_\text{eff}$ based on \eqref{eq:darcy} and \eqref{eq:eom_stress}. If $k$ and $\sigma_\text{eff}$ were functions of $\phi_v$ only, then the data from different instants in time would collapse on a single master curve. This is illustrated in figure \ref{fig:fluid_stress}b for the data plotted in figures \ref{fig:settling} and \ref{fig:fluid_stress}a. The curves for the different quantities and different times do not collapse \Rone{and the lines are highly convoluted in some regions. While this is due to the semi-logarithmic plot style as $k$ and $\sigma_\text{eff}$ approach zero, we can clearly identify three characteristic regions for both quantities: (i) a suspension region ($\phi_v<0.25$) with a high permeability, which decreases rapidly as $\phi_v$ decreases. This behavior is most pronounced at $t/\tau_s=120$, when most of the particles are still in suspension. Since there are hardly any particle-particle interactions in this region, $\sigma_\text{eff}$ approaches zero.  In addition, the lines of $k$ and $\sigma_\text{eff}$ are highly convoluted around $\phi_v \approx 0.15$. Hence, as expected, the assumption of $k$ and $\sigma_\text{eff}$ being a function of the volume fraction only does not hold in this region. (ii) A consolidating region ($0.25<\phi_v<0.58$) with an exponential decay of $k$ and increase of $\sigma_\text{eff}$, respectively, as $\phi_v$ increases. Hence, this region serves very well to derive scaling laws for $k$ and $\sigma_\text{eff}$. (iii) A jamming region ($\phi_v=0.58$), where large values for $\sigma_\text{eff}$ represent the layer formed at the very bottom of the domain that holds the weight of the entire overlying sediment. As the fluid flow ceases and the soil becomes fully consolidated, $k$ approaches zero as well.}  

As a result, we obtain a mean permeability of $k/u_s=5.7\cdot 10^{-4}$ within the sediment bed. We can convert the nondimensional permeability into dimensional quantities choosing $D_{50} = 20 \mu$m, $\rho_p =2600 \text{kg/m}^3$, $\rho_f = 1000 \text{kg/m}^3$, and $g=9.81\text{m/s}^2$. This yields $u_s = \sqrt{D_{50} g (\rho_p-\rho_f)/\rho_f}= 1.8\cdot 10^{-2}$m/s and a dimensional value of ${k=1\cdot10^{-5}}$m/s. This value corresponds very well to the permeability of a semi-pervious medium such as very fine sand or silt \citep{bear2013}, which is exactly in line with the chosen values for Co and Re that are meant to represent primary particles with a median grain size of silt.

\citet{lehir2011} and \citet{grasso2015} have used power laws as constitutive relationships to parametrize $k$ and $\langle \sigma_\text{eff}\rangle$. Based on the observation above, we can immediately conclude that fitting a power law function over all three regions will result in a poor fit. Hence, we performed a regression of a power law for regions (ii) and (iii) in figure \ref{fig:fluid_stress}b for all data that are marked by the circles. The fitted functions have different coefficients for cohesive and noncohesive sediments. A fair agreement for region (ii) can be achieved, but region (iii) is not well described for $k$. The coefficient of determination $R^2$ ranges between 0.67 and 0.81, which is satisfactory. Hence, the presented data highlights the capability of our simulation approach to improve parameterization strategies for the Gibson equation. However, larger simulations for a wider range of cohesive numbers and larger Reynolds numbers are needed to obtain a better scaling of $k$ and $\sigma_\text{eff}$.

 \setlength{\unitlength}{1cm}
\begin{figure*}
\begin{picture}(16,8)
  \put(-1.0 ,  0 ){\includegraphics[width=\textwidth,trim={0.5cm 0 0 0.5cm},clip]{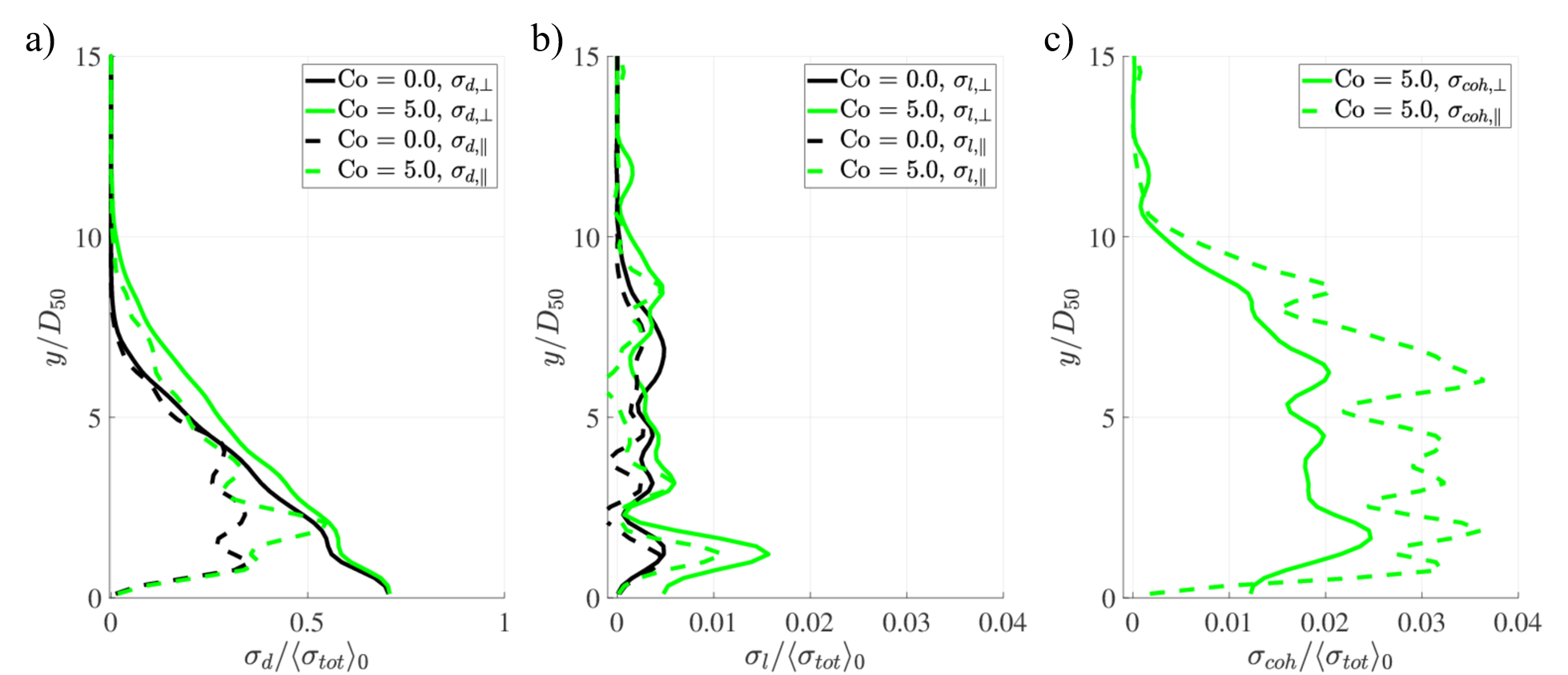}}
 %
\end{picture}
        \caption{Zoom into the horizontally averaged interparticle stress at the time when $\sigma_{c,p}/\sigma_{g,p} = 0.71$ corresponding to $t = 480\tau_s$ and $t = 346 \tau_s$ for cohesionless and cohesive sediments, respectively: (a) direct contact, (b) unresolved lubrication, and (c) cohesion.
        }
    \label{fig:intergranular_stress}
\end{figure*}

\subsection{Interparticle Stress}\label{sec:stress_internal}
While the analysis in section \ref{sec:gibson} provides insight into how the support of the granular weight transitions from hydrodynamic forces to collision forces with the bottom wall, it does not characterize internal granular stresses parallel to the wall since  sidewalls are absent in our numerical simulations. As a consequence, all interparticle forces are opposite and equal in this direction. On the other hand, the weight of the particles $\textbf{F}_{g}$ does not cancel out for contacts in the wall-normal direction so that contact forces at a given height reflect the weight of the overlying sediment. We applied operator \eqref{eq:CGM_new} to all contact forces summarized by equation \eqref{eq:particle_forces}. We denote $\sigma_\perp=\sigma_{yy}$ as the normal stress acting in $y$-direction, whereas $\sigma_\parallel= \frac{1}{2}(\sigma_{xx} + \sigma_{zz})$ denotes the normal stresses acting parallel to the wall. Having this analysis in place, we can further subdivide the effective stress into its components comprising stresses due to direct contact $\sigma_d$, unresolved lubrication $\sigma_l$, and cohesion $\sigma_\text{coh}$.

Since we focus on the sediment deposit, figures \ref{fig:intergranular_stress}a - c show a close-up of the bottom part of the domain. For cohesionless sediment, the data was taken at the final simulation stage, i.e. $t = 480 \tau_s$. At this stage, $71\%$ of the particle weight is supported by contact forces. For a meaningful comparison, we take the data from the cohesive sediment simulation at $t = 346 \tau_s$, when it has the same value of $\langle \sigma_\text{eff}\rangle_0$. For both simulations, it becomes immediately obvious that stresses due to direct contact (Figure \ref{fig:fluid_stress}a) are one order of magnitude larger than the other two components (Figure \ref{fig:fluid_stress}b and c). As expected, the nearly linear profiles of $\sigma_{d,\perp}$ reflect effective stress as the weight of the sediment beds. The steeper slope of the curve for cohesionless sediment is consistent with the fact that it is packed much more densely. The horizontal component $\sigma_{d,\parallel}$ shows significant stresses within the sediment deposit that peak at $y/D_{50} \approx 2$. These are more pronounced for cohesive grains. Throughout the sediment column, unresolved lubrication stresses remain small (Figure \ref{fig:fluid_stress}b). This holds for both simulations and is expected as most of the lubrication forces are resolved by the IBM. Figure \ref{fig:fluid_stress}c shows significant cohesive forces throughout the entire sediment deposit. Interestingly, the vertical component $\sigma_{\text{coh},\perp}$ is smaller than the horizontal component $\sigma_{\text{coh},\parallel}$. From a physical point of view, the horizontal intergranular stress due to cohesion $\sigma_{\text{coh}\parallel}$ prevents particles from arranging themselves into the densest packing formation, while the vertical cohesive stress $\sigma_{\text{coh},\perp}$ works towards a more rapid consolidation of the sediment. In the present simulation the horizontal component exceeds the vertical one, and we conclude that even small cohesive forces enable the deposited particles to remain in flocs as they are subjected to the weight of the particles settling from above. This effect accounts for the lower sediment volume fraction at the bottom of the tank for the cohesive sediment simulation (Figure \ref{fig:settling_concentration}).
%

\section{Conclusions}\label{sec:conclusions}

The present study successfully applied phase-resolved simulations to the situation of consolidation of freshly deposited cohesive and non-cohesive sediment. To this end, we have derived a stress balance based on the governing equations that makes a direct connection to the effective stress concept. The simulations fully resolve the three-dimensional flow field and provide an efficient way for excluding sidewalls from the analysis by employing periodic boundary conditions. The highly resolved data yields a physical interpretation of the effective stress as the part of the sediment weight that is supported by external contact forces with the bottom wall, as well as a full parametrization of the Gibson equation. 

An analysis of the intergranular stresses clarified the respective roles of direct contact, unresolved lubrication, and cohesive forces during consolidation and dewatering of a freshly deposited bed. For the presented data, we find that direct contact forces dominate over the other two components. As a result of the attractive Van der Waals  forces, cohesive sediment experiences larger intergranular stresses due to direct contact forces. Within the sediment deposit, cohesive forces yield intergranular stresses that lead to larger pore spaces than obtained for the consolidation of cohesionless sediment.

Hence, the first test case presented here encourages further studies with larger physical domains and a larger number of particles to achieve more realistic settling conditions. For example, larger domains could accommodate three-dimensional instabilities of particle-induced viscous fingering in the flow field  \citep{weiland1984,xu2016} and the stress balance presented here could further enhance the development of two-phase flow model closures such as the $\mu(I)$-rheology \citep{boyer2011, lee2016} or kinetic theory \citep{cheng2017}. \Rtwo{In addition, thicker sediment beds could lead to self-weight consolidation in a creeping motion \citep{sills1998,houssais2015}. To extend the present simulation approach from macroscopic silica grains towards aggregates of mud, a more realistic description of the particles is needed \cite{berlamont1993}. For example, the physical properties of the spherical particles used in the present study could be extended towards particles that are porous and compressible \cite{panah2017}, so that a description of creeping consolidation and dewatering becomes possible.}

\begin{acknowledgments}
 This research is supported by the National Science Foundation (NSF) through grant CBET-1638156 and through Army Research Office grant W911NF-18-1-0379.  BV gratefully acknowledges the Feodor-Lynen scholarship provided by the Alexander von Humboldt Foundation, Germany. The authors thank J. Israelachvili for stimulating discussions on the interaction between colloids. Computational resources for this work used the Extreme Science and Engineering Discovery Environment (XSEDE), which was supported by the National Science Foundation, USA, Grant No. TG-CTS150053. 
\end{acknowledgments}


%

\end{document}